\documentclass[11pt]{article}
\baselineskip
\pdfoutput=1
\usepackage{amsmath}
\usepackage{amssymb}
\usepackage{graphicx}
\usepackage{color}
\usepackage{cite}
\usepackage{tabls}
\usepackage{hyperref}
\usepackage{diagbox}
\usepackage{float} 
\usepackage{cancel}
\usepackage{xcolor}
\usepackage[font=small,labelfont=bf]{caption}
\newcommand{\DD}{{\rm{D}}}
\newcommand{\tmb}[1]{{\mbox{\tiny{#1}}}}

\newcommand{\eq}{\begin{equation}} 
\newcommand{\en}{\end{equation}} 
\newcommand{\eqa}{\begin{eqnarray}}
\newcommand{\ena}{\end{eqnarray}}

\newcommand{\ncl}{n_{\mathrm{cl}}}

\newcommand{\HT}{\mathrm{HT}}
\newcommand{\LT}{\mathrm{LT}}
\newcommand{\chisqred}{\chi^2_\mathrm{red}}

\newcommand{\NambuG}{Nambu-Got\={o}\;}

\begin{document}

\begin{titlepage}
\renewcommand\thefootnote{\mbox{$\fnsymbol{footnote}$}}
\begin{center}
{\Large\bf Numerical determination of the width and shape of the effective string using Stochastic Normalizing Flows}
\end{center}
\vskip1.3cm
\centerline{Michele~Caselle,\footnote{\href{mailto:michele.caselle@unito.it}{{\tt michele.caselle@unito.it}}} Elia~Cellini,\footnote{\href{mailto:elia.cellini@unito.it}{{\tt elia.cellini@unito.it}}} and Alessandro~Nada\footnote{\href{mailto:alessandro.nada@unito.it}{{\tt alessandro.nada@unito.it}}}}
\vskip1.5cm
\centerline{\sl Department of Physics, University of Turin and INFN, Turin}
\centerline{\sl Via Pietro Giuria 1, I-10125 Turin, Italy}
\vskip1.0cm

\setcounter{footnote}{0}
\renewcommand\thefootnote{\mbox{\arabic{footnote}}}
\begin{abstract}
\noindent
Flow-based architectures have recently proved to be an efficient tool for numerical simulations of Effective String Theories regularized on the lattice that otherwise cannot be efficiently sampled by standard Monte Carlo methods. In this work we use Stochastic Normalizing Flows, a state-of-the-art deep learning architecture based on non-equilibrium Monte Carlo simulations, to study different effective string models. After testing the reliability of this approach through a comparison with exact results for the Nambu-Got\={o} model, we discuss results on observables that are challenging to study analytically, such as the width of the string and the shape of the flux density. Furthermore, we perform a novel numerical study of Effective String Theories with terms beyond the Nambu-Got\={o} action, including a broader discussion on their significance for lattice gauge theories. The combination of these findings enables a quantitative description of the fine details of the confinement mechanism in different lattice gauge theories. The results presented in this work establish the reliability and feasibility of flow-based samplers for Effective String Theories and pave the way for future applications on more complex models.
\end{abstract}

\end{titlepage}

\section{Introduction}

A powerful tool to describe the infrared behaviour of pure Lattice Gauge Theories (LGTs) is Effective String Theory (EST), which models the confining flux tube joining the quark-antiquark pair as a thin vibrating string~\cite{Nambu:1974zg, Goto:1971ce, Luscher:1980ac, Luscher:1980fr, Polchinski:1991ax}. In this framework, the correlator of two Polyakov loops at a distance $R$ is associated with the full partition function of an EST:
$$\langle P(0) P^{\dagger}(R)\rangle \sim \int \DD X \; e^{-S_\tmb{EST}[X]}.$$
The simplest choice for $S_\tmb{EST}$ is the well known Nambu-Got\={o} (NG) action: this theory is known to be anomalous at the quantum level and in fact it should be considered only a large-distance approximation of the actual (anomaly free) EST. 
Despite this, the Nambu-Got\={o} action, whose partition function can be solved exactly, has been shown in the last few years to describe to high-precision the interquark potential in several different LGTs~\cite{Aharony:2013ipa, Brandt:2016xsp, Caselle:2021eir}. 
As we shall see below, the reason of this effectiveness lies in the so-called "low energy universality": even if the Nambu-Got\={o} action is only a first order approximation of the actual effective string describing the confining flux tube, the first few terms of any EST must coincide with those of the NG action. 
Exploring the terms beyond the NG action is one of the most intriguing and challenging open questions in the EST approach~\cite{EliasMiro:2019kyf, Caristo:2021tbk, Athenodorou:2011rx, Dubovsky:2014fma, Chen:2018keo, Baffigo:2023rin,Caselle:2024zoh}. 

One of the main tools to address this issue is the comparison between the results of high-precision lattice simulations and effective string predictions; however, very few EST results are known exactly. 
Besides the exact solution of the Nambu-Got\={o} partition function and a few one-loop calculations of other observables, one has to resort to approximate estimates or to educated conjectures, since analytical calculations are often too challenging.
A well-known example is the (quantum) width of the effective string, whose counterpart in gauge theories is a correlation function that measures the thickness of the confining flux tube: for this observable the only available analytical solution is a perturbative calculation up to the second order, see refs.~\cite{Gliozzi:2010jh, Gliozzi:2010zt, Gliozzi:2010zv}. 
The situation is even worse if one is interested in terms in the EST action beyond the NG term, for which even the partition function (i.e. the interquark potential measured in LGT simulations) is not known.

In a recent work~\cite{Caselle:2023mvh}, a new numerical approach has been introduced as a complementary tool to analytical EST calculations. In this approach, the effective string action is first regularized on the lattice, and then the expectation values of observables of interest are computed numerically by sampling configurations using a flow-based approach. 
Flow-based samplers~\cite{Cranmer:2023xbe} are a class of algorithms based on Normalizing Flows (NFs)~\cite{rezende2015variational}, deep generative models that provide unbiased estimators for Boltzmann-like distributions~\cite{Noe:2019, Nicoli:2019gun, Nicoli2021, Nicoli:2023qsl}. 
In ref.~\cite{Caselle:2023mvh}, the main numerical tool was a Continuous Normalizing Flow~\cite{Chen:2018keo, deHaan:2021erb, Gerdes:2022eve}. This approach worked remarkably well in the Free Boson limit of the Nambu-Got\={o} string, but it suffered from poor scaling, as larger volumes as well as smaller values of the string tension quickly became in practice out of reach. 

To address these scaling issues, in the present paper we turn to Stochastic Normalizing Flows (SNFs)~\cite{wu2020stochastic, Caselle:2022acb}: in this architecture, Normalizing Flows are combined with Non-Equilibrium Markov Chain Monte Carlo (NE-MCMC) simulations based on Jarzynski’s equality~\cite{Jarzynski1997}. The resulting algorithm, combined with a physics-informed design~\cite{Abbott:2022zsh}, can overcome the limitations of previous flow-based methods, while significantly reducing the computational cost of NE-MCMC as well. 
In particular, in this work we introduce a highly-efficient version of SNFs explicitly constructed for EST simulations, whose design is inspired by the description of the NG action as the $T\bar T$ perturbation of the free bosonic action~\cite{Caselle:2013dra, Cavaglia:2016oda}. 

First, we estimate numerically the partition function of the lattice-regularized Nambu-Got\={o} model as a benchmark of our approach, since in the NG case it is known exactly.
Besides the free energy, the main focus of our work will be the (quantum) width of the string, which is more difficult to address analytically than the partition function. Furthermore, as we shall see, this quantity is particularly sensitive to the inclusion of additional terms in the action beyond the Nambu-Got\={o} one. As a first application of our method we shall test a conjecture on the behaviour of the string width proposed a few years ago in ref.~\cite{Caselle:2010zs}.
Then, we perform an initial study of the higher-order corrections to the Nambu-Got\={o} action in EST, with a focus, again, on the string width. 
Finally, we conclude with a preliminary numerical study of the profile of the flux tube in EST through an analysis of a quartic cumulant (often referred to as the Binder cumulant) in different theoretical setups.

The work is organized as follows: in section~\ref{sec:LEST}, we introduce the lattice regularization of the Nambu-Got\={o} string, its higher order corrections and the observables of interest; then, in section~\ref{sec:SNF} we outline the numerical algorithm used to study the lattice EST. In section~\ref{sec:Results}, we present our numerical results starting from the Nambu-Got\={o} model, followed by a study of the string width for higher-order corrections and ending with an analysis of the Binder cumulant. The concluding remarks will be presented in the final section.

\section{Lattice Nambu-Got\={o} string}
\label{sec:LEST}

We report here only a few basic results on the Nambu-Got\={o} action. The interested reader can find a more detailed discussion in refs.~\cite{Aharony:2013ipa, Brandt:2016xsp, Caselle:2021eir}.
The Nambu-Got\={o} action~\cite{Nambu:1974zg,Goto:1971ce} in $D$ dimensions is defined as 
\begin{align}
\label{eq:NGaction}
S_\tmb{NG} = \sigma \int_\Sigma d^2\xi \sqrt{g},
\end{align} 
where $g\equiv \det g_{\alpha\beta}$ and
\begin{align}
\label{eq:NGaction2}
g_{\alpha\beta}=\partial_\alpha X_\mu~\partial_\beta X^\mu
\end{align} 
is the metric induced on the reference world-sheet surface $\Sigma$ by the mapping $X_\mu(\xi)$, where $\xi\equiv(\xi^0,\xi^1)$ are the world-sheet coordinates, $X_\mu$ (with $\mu=0, \dots ,D-1$) are the string coordinates in the $D$ dimensional target space and $\sigma$ is the string tension.
This action is explicitly reparametrization invariant. In the EST approach this invariance is usually fixed using the 
so-called "physical gauge", which identifies the first two degree of freedoms of the string as the world-sheet coordinates $\xi^0=X^0$, $\xi^1=X^1$. In this gauge the only remaining degrees of freedom are the transverse displacements $X^i$, with $i=2, \dots, D-1$ which are assumed to be single-valued functions of $(\xi_0,\xi_1)$. 
As it is well known this gauge fixing is anomalous, but it can be shown that the anomaly vanishes in the large $R$ limit~\cite{Aharony:2013ipa, Brandt:2016xsp} in which the Nambu-Got\={o} action can be used as an effective description of the interquark potential.  

In the following we shall set $D=3$ so as to have only one transverse degree of freedom $X(\xi^0,\xi^1)$. The generalization to larger values of $D$ is straightforward but computationally more expensive when simulated numerically.
In the physical gauge, for $D=3$, the Nambu-Got\={o} action can be rewritten as:
\begin{align}
\label{eq:NGaction3}
S_\tmb{NG}[X] = \sigma \int_0^L d\xi^0 \int_0^R d\xi^1 \sqrt{1+(\partial_{\xi^0}X)^2+(\partial_{\xi^1}X)^2},
\end{align} 

The action can be regularized on the lattice by discretizing the world-sheet on a two-dimensional square lattice $\Lambda$ with size $L\times R$ and index $x=(\tau,\epsilon)$. 
Following the usual conventions of the finite difference for the discrete derivative we obtain (see ref.~\cite{Caselle:2023mvh} for further details):
\begin{equation}
\label{eq:NG}
    S_\tmb{NG}(\phi) = \sigma \sum_{x \in \Lambda} \biggl(\sqrt{1+(\partial_{\mu}\phi(\tau,\epsilon))^2/\sigma}-1\biggr) 
\end{equation}
where
\begin{equation}
    (\partial_{\mu}\phi(x))^2 = \bigl(\phi(\tau,\epsilon)-\phi(\tau-1,\epsilon)\bigr)^2+\bigl(\phi(\tau,\epsilon)-\phi(\tau,\epsilon-1)\bigr)^2
\end{equation}
and $\phi(x)=\sqrt{\sigma}X(x)$ is a real scalar field representing the transverse degree of freedom of the string; in the context of lattice gauge theories this quantity corresponds to the density of chromoelectric flux. 

\begin{figure}[h]
  \centering
  \includegraphics[scale=0.7,keepaspectratio=true]{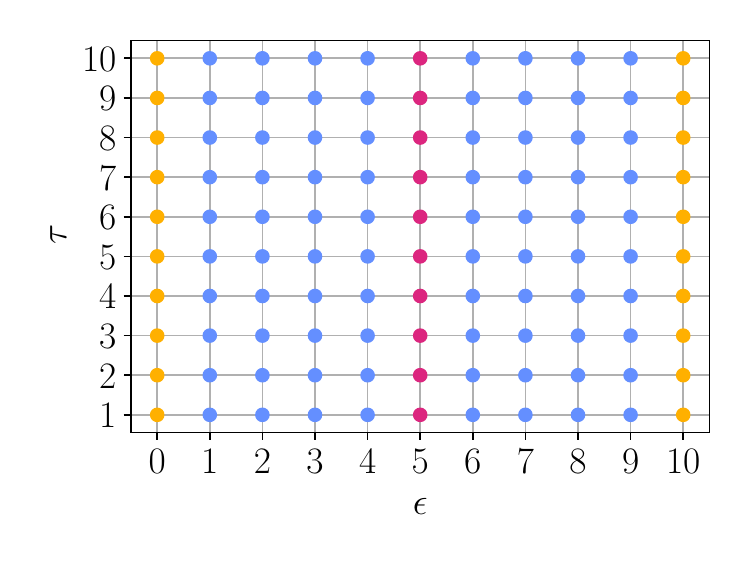}
  \caption{Schematic representation of a lattice configuration: \textcolor[HTML]{648FFF}{cyan} and \textcolor[HTML]{DC267F}{magenta} sites represent the active volume of the lattice, while \textcolor[HTML]{FFB000}{yellow} sites represent the Dirichlet boundaries where the field is fixed to $0$; the boundary in $\epsilon=R$ is considered as a part of the lattice. The width $\sigma w^2$ of eq.~\eqref{eq:LatticeWidth} is computed averaging only on the \textcolor[HTML]{DC267F}{magenta} sites.}
  \label{fig:ltconfig}
\end{figure}

While this lattice discretization holds in principle for any type of lattice, in view of the application to the description of Polyakov loop correlators in LGTs we are interested in particular on a cylinder geometry: as in ref.~\cite{Caselle:2023mvh} we fix Dirichlet boundary conditions along the $\epsilon$ direction (the lattice points at $\epsilon=R$ are considered part of the lattice volume) and periodic boundary conditions along the $\tau$ direction; see fig.~\ref{fig:ltconfig} for a schematic representation of a lattice configuration. 


\subsection{The Nambu-Got\={o} action as a $T\bar T$ perturbation}

In the physical gauge the Nambu-Got\={o} action in $D=3$ can be seen simply as an ordinary two-dimensional Quantum Field Theory (QFT) of a single bosonic degree of freedom. What is less obvious, but important for what follows, is that the NG action is actually a very peculiar case of a QFT.
In fact it can be shown~\cite{Caselle:2013dra,Cavaglia:2016oda} that it describes the irrelevant perturbation of the free two-dimensional bosonic action, driven by the $T \bar T$ operator.
The perturbing parameter is exactly the inverse of the string tension and for large values of $\sigma$ the action flows towards the theory of a Free Boson. This is easy to see if one performs a $1/\sigma$ expansion (which for dimensional reasons corresponds to a large distance expansion) of eq.~\eqref{eq:NGaction3}. 
We shall make use of this identification in the following, to construct our SNF by mimicking the $T\bar T$ flow and leveraging the fact that the Free Boson action (the starting point of the flow) can be sampled as easily as a Gaussian distribution.

In particular, in the $\sigma \to \infty$ limit, the action of eq.~\eqref{eq:NG} becomes the lattice discretization of the action of a massless scalar Free Boson (FB):
\begin{equation}
    S_\tmb{FB}(\phi)=\frac{1}{2}\sum_{x\in \Lambda}(\partial_\mu\phi)^2 ;
\end{equation}
The finite-size analytical solution for the partition function for this action can be easily computed using a Gaussian integral~\cite{Caselle:2023mvh}:
\begin{equation}
    Z_\tmb{FB}=\int D\phi e^{-S_\tmb{FB}[\phi]}=\prod_{m=1,n=1}^{L,R-1} \sqrt{ \frac{2\pi}{\lambda_{m,n} }}
\end{equation}
where, for the particular geometry in which we are interested, the eigenvalues are:
\begin{equation}\label{eq:eigenvalues}
    \lambda_k\equiv\lambda_{m,n}= 4 \sin^{2}\biggl(\frac{m\pi}{L}\biggr) + 4\sin^{2}\biggl(\frac{n\pi}{2R}\biggr).
\end{equation}
This result was the main benchmark in the test of the flow-based approach used in ref.~\cite{Caselle:2023mvh}. 
We use it again in the present work as a reference value to evaluate the contribution to the partition function of the higher order terms in the $1/\sigma$ series expansion of the Nambu-Got\={o} action: indeed, the role of these terms becomes more important as we decrease the value of $\sigma$ along the $T\bar T$ flow.

\subsection{Nambu-Got\={o} action: partition function}

A central role in EST is played by the partition function $Z$, which is directly related to the correlator of two static colour sources, i.e. to the Polyakov loop correlator in lattice gauge theory:
\begin{equation}
   \langle P(x)P^\dagger(x+R) \rangle=  Z = \int \DD X \; e^{-S_\tmb{EST}[X]}.
\end{equation}
Here $R$ is the distance between the two Polyakov loops while $L$ (the length of the lattice in the compactified direction) is the inverse temperature of the system. 

The result for the \NambuG model is known to all orders, see refs.~\cite{Luscher:2004ib, Billo:2005iv}: in three dimensions it reads
\begin{equation}
\label{eq:ZNG}
Z \sim \sqrt{\sigma} {{L}} \sum_{n=0}^{\infty} w_n K_{0}(E_n R)
\end{equation}
where $K_{0}$ is the modified Bessel function of the second kind of order $0$, $w_n$ the multiplicities of the closed string states that propagate from one Polyakov loop to the other and the $E_n$ their energies:
\begin{equation}
 \label{eq:En}
  {E}_n(L)=\sigma L \sqrt{1 + \frac{8\pi}{\sigma L^2} \left( n-\frac{1}{24}\right)}.
\end{equation}

This expression is valid for any choice of $L$ and $R$, but it simplifies a lot in the large-$R$ limit in which the sum  
is dominated by the lowest state $n=0$. In this limit we can approximate the Bessel function to the leading term of its large distance expansion: $K_0(x)\sim \exp(-x)/\sqrt{x}$.
Since $w_0=1$ we end up with this simple expression
\begin{equation}
\label{eq:logZNG}
    -\log Z = \frac{1}{2} \log\frac{R}{L} + \sigma (L) RL + C
\end{equation}
where the constant $C$ has no further dependence on $R$ or $L$ and we introduced the nonzero-temperature string tension $\sigma(L)$ as a short-hand for
\begin{equation}
\label{eq:SigmaL}
    \sigma(L) =\sigma \sqrt{1 - \frac{\pi}{3 \sigma L^2}}.
\end{equation}

Interestingly, this limit corresponds to the high-temperature limit of the interquark potential in which we choose a temperature $1/L$ still in the confined phase, but large enough to fulfill the constraint $R\gg L$. 
Eq.~\eqref{eq:SigmaL} can be interpreted in the LGT framework as describing the temperature dependence of the string tension. As we increase the temperature the string tension decreases: when the argument of the square root finally vanishes for $L_c=\sqrt{\pi /3\sigma}$ we reach the deconfinement transition.
This analytical solution represents the main benchmark we will use to verify the reliability of the numerical approach described in section~\ref{sec:SNF}, in which the fine details of the \NambuG action are explored numerically.

\subsection{Nambu-Got\={o} action: width and shape of the flux tube}
\label{sec:newNGwidth}

The second observable of interest in this work is the width of the string, which can be computed in the lattice version of the theory as
\begin{equation}
\label{eq:LatticeWidth}
    \sigma w^2 (\sigma, L, R) = \langle \phi^2(\tau,R/2)\rangle_\tau.
\end{equation}
As in ref.~\cite{Caselle:2023mvh} the $\langle \dots \rangle_\tau$ expectation value takes the average only over the temporal extension, and the scalar fields $\phi$ are fixed to be at the half-distance $R/2$ between the colour sources, see fig.~\ref{fig:ltconfig}.

The only known analytical expression for the width in EST is a perturbative calculation up to the second order in $1/\sigma$, see refs.~\cite{Gliozzi:2010jh, Gliozzi:2010zt, Gliozzi:2010zv}. 
In the low-temperature regime $L\gg R$ we have that
\begin{equation}
 \sigma w^2(\sigma, L, R) = \frac{1}{2\pi} \log \frac{R}{R_c} \left( 1 - \frac{\pi}{4\sigma R^2} \right)+ \frac{5}{96} \frac{1}{\sigma R^2} + \dots
\label{eq:w2FBLT}
\end{equation}
where $R_c$ is a new scale of the model, which emerges from the regularization of the correlator defining the string width.
This scale sets a threshold below which the EST picture cannot be trusted anymore and in LGTs applications it is associated to the so called "intrinsic width" of the flux tube (see ref.~\cite{Caselle:2012rp} for a discussion of this issue). We will comment further on the role of this scale in our case later in this contribution.

In the high-temperature regime $R \gg L$ the expected behaviour is
\begin{equation}
\label{eq:w2FBHT}
 \sigma w^2(\sigma, L, R) = \frac{1}{2\pi} \log \frac{L}{L_c} + \frac{R}{4L} + \frac{\pi}{24}\frac{R} {\sigma L^3}+\cdots
\end{equation}
where the new scale $L_c$ has the same origin of $R_c$ but, interestingly, it does not affect the $R$ dependence of the string width. 

In the high-temperature limit it has been conjectured, using the Svetitsky-Yaffe mapping of the LGT into a suitable two-dimensional spin model, that the terms multiplying the $R/4L$ factor can be resummed~\cite{Caselle:2010zs}:
$$\biggl(1+\frac{\pi}{6\sigma L^2}+\cdots\biggr) = \frac{1}{\sqrt{1 -  \frac{\pi}{3 \sigma L^2}}} = \frac{\sigma}{\sigma(L)}.$$
Thus, the behaviour of the linear term in $R$ of the Nambu-Got\={o} string width is expected to be:
\begin{equation}
\label{eq:w2NGHT}
    w^2(\sigma, L, R) = \frac{1}{\sigma(L)}\frac{R}{4L} + \cdots
\end{equation}
A reliable numerical test of this conjecture is one of the main goals of this work.

The numerical approach described in the next section allows in principle for the calculation of any observable over the probability density of the EST model under study. Thus, we focus our attention also on a new observable defined as a combination of the quartic and quadratic moments of the flux density in the midpoint $\phi(\tau,R/2)$:
\begin{equation}
\label{eq:binder}
    U=1-\frac{\langle \phi^4(\tau,R/2)\rangle_{\tau}}{3\langle \phi^2(\tau,R/2)\rangle_{\tau}^2}.
\end{equation}
As in eq.~\eqref{eq:LatticeWidth}, the expectation values $\langle \dots \rangle_{\tau}$ are computed fixing the spatial coordinate to $R/2$. In the literature (albeit in different contexts) the quantity $U$ is called the Binder cumulant and we shall denote it as such in the following. It is identically zero if $\phi(\tau,R/2)$ follows a Gaussian distribution: in this sense, it can be used as a probe of the non-Gaussianity of the flux density. In the LGT framework it is used to describe the "shape" of the flux tube, see ref.~\cite{Verzichelli:2025cqc}. 

Finally, to avoid confusion, let us stress that the string width that we discuss in this section is a purely quantum effect, as at the classical level the Nambu-Got\={o} string has zero width. This is one of the reasons behind the difficulty in obtaining analytical  results for this quantity. 

\subsection{Higher-order terms beyond the Nambu-Got\={o} action}

The study of the higher-order corrections beyond the \NambuG action represents one of the most interesting open problems in EST. 
These corrections must respect Poincar\'e invariance in the target space and thus the first few allowed terms for a three-dimensional target space are\footnote{In four dimensions, due to the presence of two transverse fields two different invariants can be constructed at level four.}
\begin{align}
    S_\tmb{EST} =  \int_\Sigma d^2\xi \sqrt{g}\bigl[\sigma+\gamma_1 \mathcal{R}+\gamma_2 \mathcal{K}^2+\gamma_3 \mathcal{K}^4\cdots\bigr]
\end{align}
where $\gamma_i$ are coupling constants, $\mathcal{R}$ represent the Ricci scalar, and $\mathcal{K}$ is the extrinsic curvature. 
This expansion is strongly constrained by the so called "low energy universality" which states that the first two terms can be neglected when studying ordinary non-Abelian LGTs. 
In fact $\mathcal{R}$ is a topological invariant in two dimensions and it can be neglected.
Furthermore, the $\mathcal{K}^2$ term is proportional to the equation of motion of the Nambu-Got\={o} action and can be eliminated by a suitable field redefinition. There are however a few exceptions to the last statement: the most interesting one for our purposes is represented by the Polchinski-Yang solution of the rigid string~\cite{Polchinski:1992ty}, in which the Nambu-Got\={o} term is treated as a small perturbation of the $\mathcal{K}^2$ one. This particular case seems to be realized in the so called "reconfined phase" of trace deformed Lattice Gauge Theories~\cite{Unsal:2008ch}. For this reason we studied (separately) both the inclusion of the $\mathcal{K}^2$ and of the $\mathcal{K}^4$ term in the EST action looking at the following two "Beyond \NambuG" (BNG) actions, namely:
\begin{align}
    S^1_\tmb{BNG} &= S_\tmb{NG}+S_\tmb{$\mathcal{K}^2$} \label{eq:NGK2action}\\
    S^2_\tmb{BNG} &= S_\tmb{NG}+S_\tmb{$\mathcal{K}^4$} \label{eq:NGK4action}.
\end{align}
It is interesting to notice that the first of these two actions has indeed a long history. Originally introduced to describe the physics of fluid membranes~\cite{Peliti:1985eo, Helfrich:1985eo, Forster:1986ot}, it was later proposed by Polyakov and by Kleinert as a way to stabilize the Nambu-Got\={o} action~\cite{Polyakov:1986cs, Kleinert:1986bk}. 
It is often denoted as "rigid string" since the $\mathcal{K}^2$ term is expected to increase the stiffness of the string. As we shall see below, our results on the string width confirm this intuition. To the best of our knowledge, this is the first time that an explicit quantitative evidence of the effect of the rigidity term in the width of the rigid string is reported. 

Following ref.~\cite{EliasMiro:2019kyf} we approximate $\mathcal{K}^2$  to the first order:
\begin{equation}
\label{eq:LK2}
    \mathcal{K}^2 \sim \bigl((\partial_0\partial_0 \phi(x))^2+(\partial_1\partial_1 \phi(x))^2+2(\partial_1\partial_0 \phi(x))^2 \bigr).
\end{equation}
Thus, we can write the discretized version of the two terms as follows:
\begin{align}
    S_\tmb{$\mathcal{K}^2$}(\phi) &=\gamma_2 \sum_{x \in \Lambda} \mathcal{L}_\tmb{$\mathcal{K}^2$}(\phi(x)) \label{eq:K2} \\
    S_\tmb{$\mathcal{K}^4$}(\phi) &=\gamma_3 \sum_{x \in \Lambda} \bigl(\mathcal{L}_\tmb{$\mathcal{K}^2$}(\phi(x))\bigr)^2 \label{eq:K4}
\end{align}
where:
\begin{equation}
\begin{split}
   \mathcal{L}_\tmb{$\mathcal{K}^2$}(\phi(\tau,\epsilon))&=\bigl(\phi(\tau+1,\epsilon)-2\phi(\tau,\epsilon)+\phi(\tau-1,\epsilon)\bigr)^2+\\
   &+\bigl(\phi(\tau,\epsilon+1)-2\phi(\tau,\epsilon)+\phi(\tau,\epsilon-1)\bigr)^2+\\
   &+\frac{1}{8}\bigl(\phi(\tau+1,\epsilon+1)+\phi(\tau-1,\epsilon-1)-\phi(\tau+1,\epsilon-1)-\phi(\tau-1,\epsilon+1)\bigr)^2
\end{split}
\end{equation}
Some special care has to be taken on the boundary $\epsilon=R$: the regularization must take into account the Dirichlet boundary conditions. The non-vanishing terms are:
\begin{equation}
    \mathcal{L}_\tmb{$\mathcal{K}^2$}(\phi(\tau,R))=\bigl(\phi(\tau,\epsilon-2)-2\phi(\tau,\epsilon-1)\bigr)^2+\frac{1}{2}\bigl(\phi(\tau-1,\epsilon-1)-\phi(\tau+1,\epsilon-1)\bigr)^2
\end{equation}
In the following we study numerically the actions of eqs.~\eqref{eq:K2} and \eqref{eq:K4}, by keeping $\sigma$ fixed and using the NG action as the starting point of our flow-based approach. The perturbing parameters of the flows will be $\gamma_2$ and $\gamma_3$ respectively. Our goal will be to understand how the shape and thickness of flux tube change as a consequence of the addition of the $\mathcal{K}^2$ or $\mathcal{K}^4$ terms.

\section{Sampling the Nambu-Got\={o} action with Stochastic Normalizing Flows}
\label{sec:SNF}

Stochastic Normalizing Flows (SNFs)~\cite{wu2020stochastic,Caselle:2022acb} are a class of generative models that combine non-equilibrium Markov Chain Monte Carlo (NE-MCMC) simulations based on Jarzynski's equality~\cite{Jarzynski1997}, that have seen widespread use in lattice field theory in recent years (see refs.~\cite{Caselle:2016wsw, CaselleSU3, Francesconi:2020fgi, Bulgarelli:2023ofi, Bonanno:2024udh,Bulgarelli:2024onj}), with Normalizing Flows, a deep-learning approach that we briefly discuss later in this section.

Jarzynski's equality is a powerful relation in non-equilibrium statistical mechanics that relates the ratio of the partition functions of two equilibrium distributions to an average over out-of-equilibrium processes:
\begin{equation}
    \label{eq:JE}
    \frac{Z}{Z_0}=\langle \exp(-w(\phi_0,\phi_1,...,\phi_N))\rangle_\mathrm{f}.
\end{equation}
Each non-equilibrium evolution starts from a configuration sampled from the initial "prior" distribution 
$$q_0(\phi_0) = e^{-S_0(\phi_0)}/Z_0$$ 
and ends after $N$ steps at the final "target" distribution 
$$p(\phi)=e^{-S(\phi)}/Z.$$ 
During the evolution the intermediate configurations $\phi_n$ are sampled with a Monte Carlo algorithm that satisfies detailed balance using the intermediate action $S_{\eta_n}$ defined by a protocol $\eta_n$. The protocol represents the set of parameters that interpolate the prior action $S_{\eta_{0}}\equiv S_0$ and the target action $S_{\eta_{N}} \equiv S$. 
The quantity $w(\phi_0,\phi_1,...,\phi_N)$ is the dimensionless work done on the system during each evolution, defined as:
\begin{equation}
    \begin{split}
        w(\phi_0,\phi_1,...,\phi_N)  &= S_{\eta_N}[\phi_N]-S_{\eta_0}[\phi_0]-Q(\phi_0,\phi_1,....,\phi_N) \\
        Q(\phi_0,\phi_1,....,\phi_N) &= \sum_{n=0}^{N-1}\{ S_{\eta_{n+1}}[\phi_{n+1}]-S_{\eta_{n+1}}[\phi_{n}] \}
    \end{split}
\end{equation}
where we also defined the heat $Q$.
In eq.~\eqref{eq:JE} the average $\langle ... \rangle_f$ is taken over all the possible realizations of the evolution defined by a protocol $\eta_n$.
We refer for example to refs.~\cite{Caselle:2016wsw,CaselleSU3} for a derivation of these results in the context of Markov Chain Monte Carlo for lattice field theory.
Let us stress that only the configurations at the beginning of each evolution must be at thermodynamic equilibrium, i.e. sampled from the prior distribution $q_0$. Conversely, at each intermediate step the configuration $\phi_n$ will be generated using an action $S_{\eta_n}$, which will in general drive the system away from the equilibrium in a way dictated by the protocol $\eta_n$.
Finally, computing the expectation value of general observables is possible using a reweighting-like formula similar to eq.~\eqref{eq:JE}, namely
\begin{equation}
\label{eq:obsNE}   
     \langle \mathcal{O} \rangle_p =\frac{\langle \mathcal{O}(\phi_N) \exp(-w(\phi_0,\phi_1,...,\phi_N))\rangle_f}{\langle \exp(-w(\phi_0,\phi_1,...,\phi_N))\rangle_f},
\end{equation}
where $\langle \dots \rangle_p$ is the expectation value according to the target distribution $p$; we refer to ref.~\cite{Bonanno:2024udh} for further details on out-of-equilibrium evolutions.

Non-equilibrium Markov Chain Monte Carlo can be combined with Normalizing Flows~\cite{rezende2015variational} to build SNFs. 
A Normalizing Flow is a parametric diffeomorphism $g_\theta: \mathbb{R}^d\to\mathbb{R}^d$ which maps a prior $q_0$ into a learned distribution $q_\theta$ that approximates the target $p(\phi)=\exp(-S[\phi])/Z$. 
Normalizing Flows are generally built as a discrete sequence of transformations, i.e.
$$g_\theta=g_{\theta_{\ncl}}\circ ...\circ g_{\theta_i}\circ ... \circ g_{\theta_1}$$
where $\ncl$ is the total number of the invertible transformations $g_{\theta_i}$, called coupling layers. The transformation of each layer $i$ is implemented using neural networks with parameters $\theta_i$. 
Normalizing Flows make the exact computation of the density of the samples possible using the change of variable theorem: in particular, the probability density of the output $g_{\theta} (\phi_0)$ of a Normalizing Flow can be written as: 
\begin{equation}\label{eq:logDet}
 \begin{split}
     q_{\theta}(g_{\theta}(\phi_0)) &= q_0(\phi_0) \, e^{- \log J} \\
     \log J &=\sum_{n=0}^{\ncl} \log |\det J_n(\phi_n)|
 \end{split}
\end{equation}
where $J_n(\phi_n)$ is the Jacobian of the field variable transformation performed in the $n$-th layer of the flow.
 
Finally, Stochastic Normalizing Flows are created by interleaving non-equilibrium Monte Carlo updates with Normalizing Flow layers. The (generalized) work is now defined as 
\begin{equation}
 w(\phi_0,\phi_1,...,\phi_N)  = S_{\eta_N}[\phi_N]-S_{\eta_0}[\phi_0] -Q - \log J 
\end{equation}
where the term $Q$ comes from the $N$ stochastic updates and the term $J$ from the Jacobian of the $\ncl$ coupling layers. 
The training of SNFs is performed by minimizing the Kullback-Leibler divergence between the probability of "forward" (from prior to target) and "reverse" evolutions, namely
\begin{equation}\label{eq:DKL}
    D_{KL}(q_0P_f||q_NP_r)=\langle w(\phi_0,\phi_1,...,\phi_N) \rangle_f + \log{\frac{Z_N}{Z_0}}
\end{equation}
where $P_f$ and $P_r$ represent respectively the forward and the reverse transition probabilities of each SNFs: we refer to~\cite{Caselle:2022acb} for further details. 

In this paper, we took inspiration from the $T\bar T$ irrelevant integrable perturbations~\cite{Cavaglia:2016oda, Smirnov:2016lqw} to design our Physics-Informed Stochastic Normalizing Flows (PI-SNF). Specifically, we used as a prior distribution a massless free field regularized on the lattice:
\begin{equation}
\label{eq:FB}
    S_{\eta_{0}}[\phi] = S_0[\phi] =\frac{1}{2}\sum_{x\in \Lambda}(\partial_\mu\phi)^2,
\end{equation}
while we chose the intermediate actions for the non-equilibrium Monte Carlo updates to correspond simply to the Nambu-Got\={o} action with a string tension $\sigma_n$, i.e.
\begin{equation}
\begin{split}
    q_{\eta_{n}}[\phi] & = \exp(-S_{\eta_{n}}[\phi])/Z_{n}\\
    S_{\eta_{n}}[\phi] & = \sigma_n \sum_{x \in \Lambda} \biggl(\sqrt{1+(\partial_{\mu}\phi(x))^2/\sigma_n}-1\biggr).
\end{split}
\end{equation}
The specific protocol for $\sigma_n$ that we have followed in this work is a linear interpolation in the inverse of the string tension, starting from $1/\sigma=0$ (i.e. the action of eq.~\eqref{eq:FB}) and ending at the inverse of the target value of the string tension. In the case of the higher-order corrections we use as a prior distribution the NG theory (again obtained with a SNF) and then the couplings $\gamma_2$ or $\gamma_3$ are switched on using a linear protocol. 

\section{Numerical Results}\label{sec:Results}

\subsection{Simulation details}

The PI-SNFs used in this work are made of "blocks", each of which is composed by two affine coupling layers with even-odd masks~\cite{Dinh:2017} and one Hybrid Monte Carlo (HMC)~\cite{DUANE1987216} update. 
The two affine layers in each block share the same convolutional neural networks, while the number of kernels and of hidden layers varies depending on the lattices under study. 
The HMC update uses a leapfrog integrator with 10 steps and a stepsize of $0.1$. 
The free boson distribution (representing the prior distribution in our method) is sampled with a procedure similar to the one used in ref.~\cite{Abbott:2022zsh}: we first sample a Gaussian distribution with identity covariance, and then we rescale the samples using the eigenvalues of the theory to obtain free field configurations in the momentum space.
Finally, we transform the samples using the eigenfunctions of the analytical solution: the eigenvalues and eigenfunctions of the free field solution are discussed in ref.~\cite{Caselle:2023mvh}. 

To train the models, we strongly rely on transferring the parameters of the networks between volumes. We first trained a single PI-SNF for a $40\times 40$ lattice at fixed target coupling for $5000$ iterations of the Adam algorithm~\cite{Kingma:2014vow} with batch size $32$ and learning rate $0.0001$; we then used the weights of this flow as the initialization for the other lattice volumes and trained for $100$ iterations with learning rate $0.00001$.  

The partition function and the expectation value of the width are computed using eqs.~\eqref{eq:JE} and \eqref{eq:obsNE} respectively. In order to study the Binder cumulant we used the output of the PI-SNFs as a proposal for an independent Metropolis-Hasting procedure, see for example ref.~\cite{Albergo1}: the expectation values of eq.~\eqref{eq:binder} are then computed on the configurations of the new Markov Chain.
The code is based on the PyTorch library~\cite{Pythorc} and the statistical analysis is performed using pyerrors~\cite{JOSWIG2023108750}. 

\subsection{Nambu-Got\={o}: partition function}
\label{sec:resultsNG}

We start by using analytical results for the Nambu-Got{\=o} partition function as a benchmark of our approach. 
In particular, our goal is to test the ability to clearly distinguish the prediction of the NG action from that of the free bosonic action, i.e if we are able to see the effect of the higher order terms in the NG action beyond the first order approximation. 

In order to pursue this program, we had to reach small values of the string tension $\sigma$ while keeping the lattice sizes reasonably large. 
This is a particularly demanding setting even for flow-based approaches, as the CNF used in ref.~\cite{Caselle:2023mvh} could sample only values of $\sigma \sim \mathcal{O}(10)$. 
The SNF architecture used in this work represents a substantial improvement in the scaling with the string tension, as values of $\sigma \sim \mathcal{O}(10^{-2})$ are now accessible\footnote{We stress that in ref.~\cite{Caselle:2023mvh} it was showed that CNFs scaled much better than a standard HMC simulation and were by far the more efficient sampler.}.

In order to make the analysis more straightforward, we set our simulations in the high-temperature regime in which the expected behaviour of the NG free energy is given by eq.~\eqref{eq:logZNG}.
We computed the partition function of the NG model for $\sigma=1/30$ and $\sigma=1/50$ and for lattices with $R \in [60,100]$ and $L \in [10,15]$.
We first fitted the results for $-\log Z$ in $R$ with an expression inspired by eq.~\eqref{eq:logZNG}, namely
\begin{equation}
\label{eq:NGlnZ1}
    -\log Z= a(L) \sigma R + b(L) + c(L) \log(R).
\end{equation}
The results for the various coefficients are reported in table~\ref{tab:logZNG}. 

\begin{figure}[t]
  \centering
  \includegraphics[scale=0.6,keepaspectratio=true]{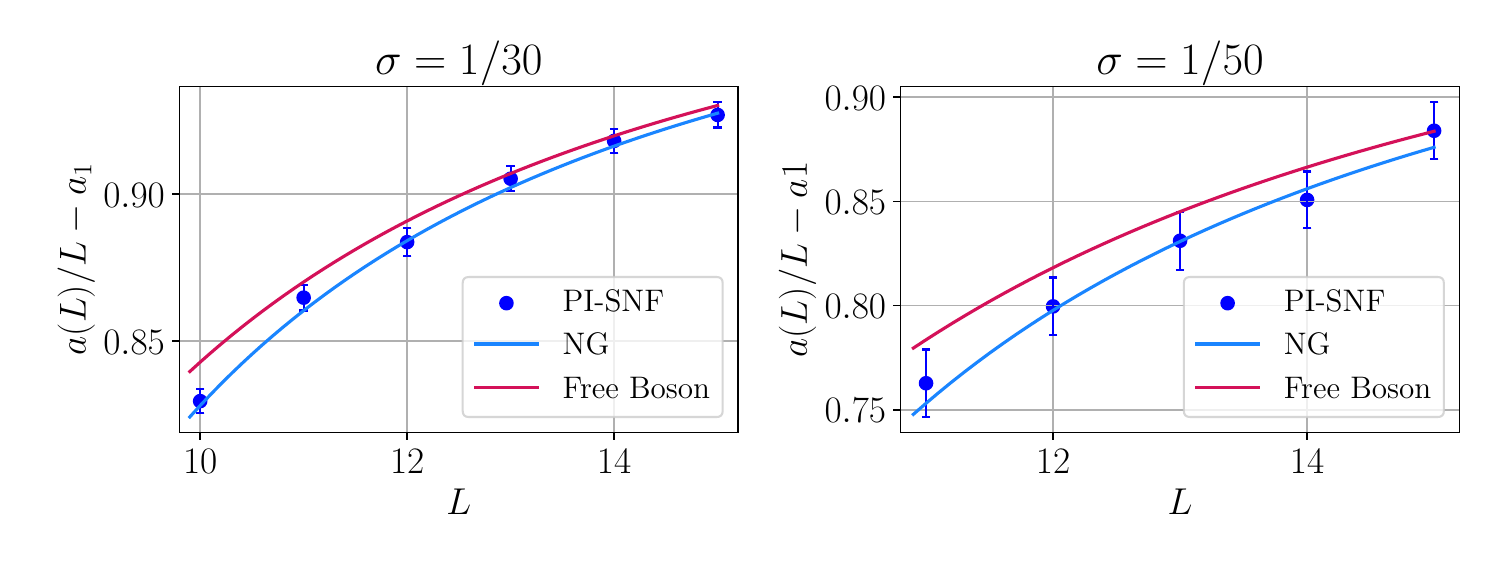}
  \caption{Results for the coefficient $a(L)$ of the partition function of the lattice Nambu-Got\={o} model, for various values of $L$ and for $\sigma=1/30$ (left panel) and $\sigma=1/50$ (right panel). The expected behaviour for the Free Boson and the Nambu-Got\={o} actions are also shown.}
  \label{fig:logZNG}
\end{figure}

Next, we take a closer look at the coefficient $a(L)$ by using a functional form reminiscent of the NG behaviour of eq.~\eqref{eq:SigmaL}. In particular, we fit $a(L)$ in $L$ using
\begin{equation}
\label{eq:NGlnZ2}
    a(L) = \left( \sqrt{1-\frac{a_0}{\sigma L^2}} + a_1 \right) \, L
\end{equation}
where $a_1$ is an undetermined bulk constant, while according to eq.~\eqref{eq:logZNG} we expect $a_0=\frac{\pi}{3}= 1.047 \dots$.
Results of the fits are reported in table~\ref{tab:logZNGa} and the corresponding curves are displayed in fig.~\ref{fig:logZNG}.

\begin{table}[H]
   \centering
   \begin{tabular}{|c | c c c|} 
     \hline
     $ \sigma$ & $a_0$ & $a_1$ & $\chisqred$ \\ [0.5ex]
     \hline
     $1/30$ & $1.03(2)$ & $-52.517(3)$ & $0.33$ \\
    \hline\hline
     $1/50$ & $1.04(7)$ & $-98.99(1)$ & $1.56$ \\
    \hline\hline
  \end{tabular}
   \caption{Results for the coefficients of the fit of $a(L)$ of eq.~\eqref{eq:NGlnZ2}.}
   \label{tab:logZNGa}
\end{table}

We find in both fits an excellent agreement of the parameter $a_0$ with the expected result $\frac{\pi}{3}= 1.047 \dots$. Crucially, the difference between the whole square root behaviour of eq.~\eqref{eq:NGlnZ2}, a clear fingerprint of the NG action, and the free bosonic action is clearly visible in our data (see fig.~\ref{fig:logZNG}). This provides compelling evidence that other observables can be reliably and efficiently addressed using a SNF architecture.
 
\subsection{Nambu-Got\={o}: string width}

The natural next step in this program is to use numerical simulations to compute observables that lack an analytical description. 
In particular, as we mentioned above, we are interested in the string width $\sigma w^2$ (see eq.~\eqref{eq:LatticeWidth}): we performed independent simulations with $R \in [30,100]$ and $L \in [5,20]$ at fixed $\sigma=1/10$.

We focus our analysis on the high temperature ($R\gg L$) setting; mimicking the behaviour of eq.~\eqref{eq:w2NGHT} we fitted the $R$ dependence of our results with the following expression:
\begin{equation}
\label{eq:width1}
    \sigma w^2(L,R) = f(L) R + g(L);
\end{equation}
the results of the two fit parameters are reported in table~\ref{tab:widthNG}.
Then, we fit again in $L$ the coefficient $f(L)$ using
\begin{equation}
\label{eq:width2}
    f(L)= \frac{1}{4L} \left( \frac{1}{\sqrt{1 - \frac{f_0}{\sigma L^2}}} + f_1 \right)
\end{equation}
where $f_1$ is an undetermined bulk constant and $f_0$, according to the conjecture mentioned in section~\ref{sec:newNGwidth}, should take the value $\pi/3$.
Results of the fit are reported in table~\ref{tab:widthNGf}.

\begin{table}[h]
    \centering
\begin{tabular}{|c c c|} 
 \hline
 $f_0$ & $f_1$ & $\chisqred$ \\ [0.5ex]
 \hline\hline
 $1.01(9)$ & $12.35(1)$ & $0.47$ \\
 \hline
\end{tabular}
    \caption{Results for the coefficients of the fit of $f(L)$ of eq.~\eqref{eq:width2}. }
    \label{tab:widthNGf}
\end{table}

\begin{figure}[h]
  \centering
  \includegraphics[scale=0.65,keepaspectratio=true]{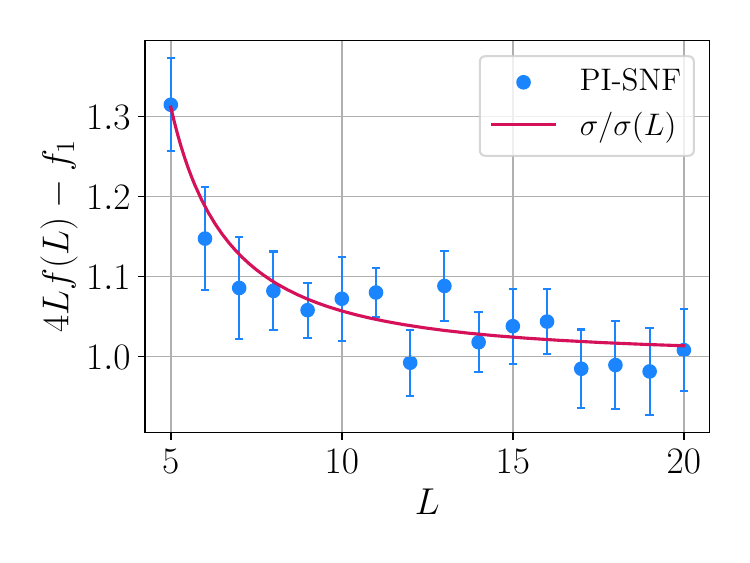}
  \caption{Results for the numerical prediction of the term $\sigma/\sigma(L)$, which corresponds to the coefficient $f(L)$ of the string width for $\sigma = 1/10$, after removing the divergent term $f_1$ and multiplying by $4L$. The solid curve corresponds to the true value from eq.~\eqref{eq:SigmaL}.}
  \label{fig:W2NG}
\end{figure}

Looking at table~\ref{tab:widthNGf} we see that our result $f_0=1.01(9)$ remarkably agrees with the conjectured coefficient $\pi/3=1.047...$. This can be appreciated in fig.~\ref{fig:W2NG} as well, where we compare our data for the combination $4Lf(L)-f_1$, which corresponds to the numerical prediction for the term $\sigma/\sigma(L)$, with the true value reported in eq.~\eqref{eq:SigmaL}. Notice that the free bosonic solution would correspond to a flat horizontal line, which is clearly excluded by the data.

\subsection{Beyond Nambu-Got\={o}: the string width in presence of the $\mathcal{K}^2$ and $\mathcal{K}^4$ terms}
\label{sec:resultsK2K4}

\begin{figure}[h]
  \centering
  \includegraphics[scale=0.44,keepaspectratio=true]{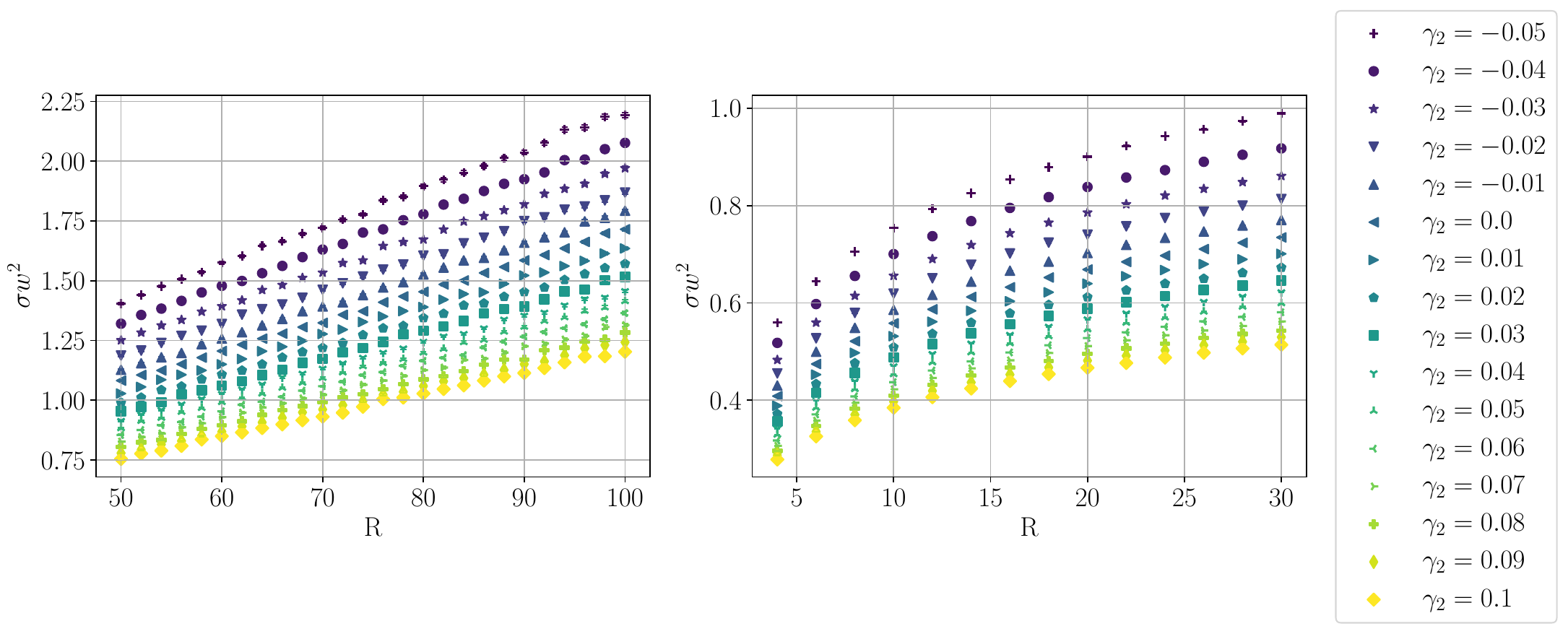}%
  \caption{Results for the width $\sigma w^2(R,L,\gamma_2)$ at high temperature ($R \gg L=20$, left panel) and at low temperature ($R \ll L=80$, right panel). The $\gamma_2=0$ case is equivalent to the Nambu-Got\={o} action.}
  \label{fig:wK2}
\end{figure}

We performed numerical simulations in the presence of a $\mathcal{K}^2$ term both in the high- ($L\ll R$) and low- ($L \gg R$) temperature limits, where the theoretical expectations for the free bosonic string have simple analytical expressions, see eqs.~\eqref{eq:w2FBLT} and \eqref{eq:w2FBHT}. 
We set $\gamma_2 \in [-0.05,0.1]$, we fixed $\sigma=100$: all results for the width are plotted in fig.~\ref{fig:wK2} for both regimes.

As it can be seen from both panels of fig.~\ref{fig:wK2} the rigidity term has a strong effect on the string width. 
As $\gamma_2$ increases the width of the string decreases; its dependence on $R$ changes as well and it becomes flatter.  
This is consistent with the qualitative picture in which the rigidity term "squeezes" the quantum fluctuations of the string: yet, the magnitude of this change is surprising. 
We made an attempt to analyse the contribution of the rigidity term by fitting our data with a functional form inspired by the free boson behaviour of eqs.~\eqref{eq:w2FBLT} and \eqref{eq:w2FBHT}. 

\begin{figure}[h]
  \centering
  \includegraphics[scale=0.6,keepaspectratio=true]{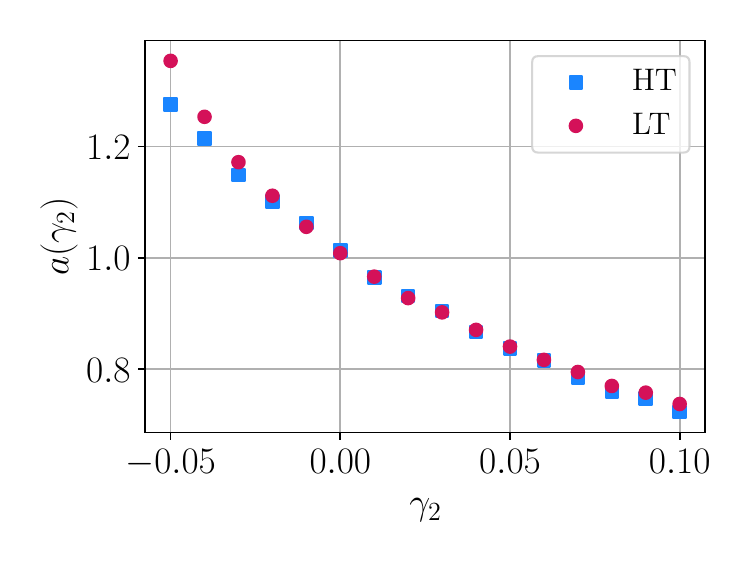}
  \includegraphics[scale=0.6,keepaspectratio=true]{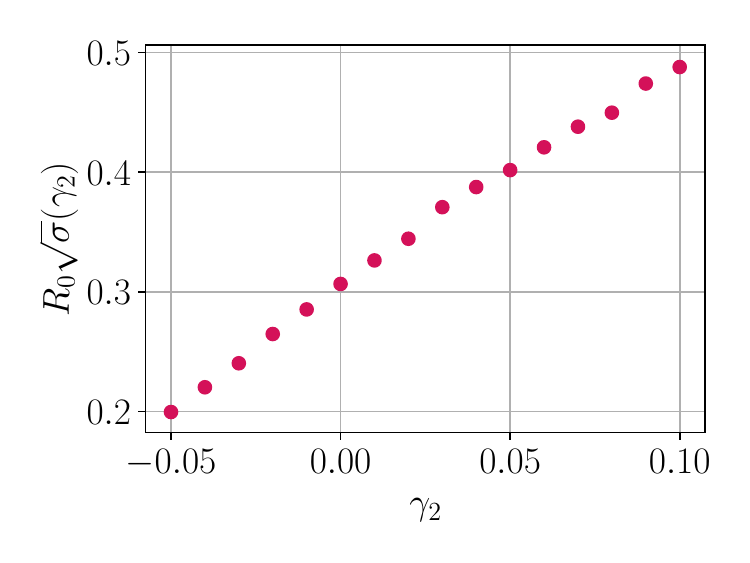}
  \caption{Results for $a^{(\HT)}(\gamma_2)$ and $a^{(\LT)}(\gamma_2)$ (left panel), and for $R_c(\gamma_2)$ (right panel).}
  \label{fig:aRgamma2}
\end{figure}

Studying the high-temperature regime first, we fixed $L=20$ and chose values of the spatial extent so that $R\gg L$; then we fitted the $R$ dependence of our results using the following functional form
\begin{equation}
\label{eq:wK2HT}
    \sigma w^2_{(\HT)} (R,L,\gamma_2) = a^{(\HT)}(\gamma_2) \frac{R}{4L} + b^{(\HT)}(\gamma_2)
\end{equation}
This two-parameter fit works extremely well for all values of $\gamma_2$: the results are reported in table~\ref{tab:K2HT} and the parameter $a^{(\HT)}(\gamma_2)$ is displayed in fig.~\ref{fig:aRgamma2} as well. 
The $\gamma_2=0$ result for $a^{(\HT)}$ is quite close to $1$, which is the value expected for the Free Boson action and consistent with the large value of $\sigma$. 
More interestingly, deviations from this value at $\gamma_2 \neq 0$ are quite remarkable and clearly visible.


The same happens in the low-temperature regime, which we realized by fixing $L=80$ and choosing this time $R\ll L$. In this case we fitted the data with the following functional form inspired by eq.~\eqref{eq:w2FBLT}:
\begin{equation}
\label{eq:wK2LT}
    \sigma w^2_{(\LT)} (R,L,\gamma_2) = a^{(\LT)}(\gamma_2) \frac{\log R}{2\pi} + b^{(\LT)}(\gamma_2) + c^{(\LT)}(\gamma_2) \frac{1}{R^2}
\end{equation}
Results for the various coefficients are reported in table~\ref{tab:K2LT}, with $a^{(\LT)}(\gamma_2)$ being plotted in the left panel of fig.~\ref{fig:aRgamma2} as well. Interestingly, the behaviour of the coefficient of the logarithm looks qualitatively similar to that of $a^{(\HT)}(\gamma_2)$ in the high-temperature limit.

Finally, we can relate the $R_c(\gamma_2)$ term as it appears in eq.~\eqref{eq:w2FBLT} with the $b^{(\LT)}(\gamma_2)$ coefficient simply assuming
\begin{equation}
\label{eq:K2R0b}
    R_c(\gamma_2)=\exp{\bigl(-2\pi b^{(\LT)}(\gamma_2)\bigr)};
\end{equation}
we show the values of this term in the right panel of fig.~\ref{fig:aRgamma2}. 

Interestingly, all the values of $R_c$ (and in particular the one at $\gamma_2=0$ which corresponds to the pure Nambu-Got\={o} model) are below one lattice spacing, providing evidence that we can trust our model down to very small values of $R$ or $L$.  
Furthermore, a value of $R_c$ smaller than one lattice spacing is perfectly consistent with the absence in our model of an intrinsic width of the flux tube.
Conversely, the value of $R_c$ measured in non-Abelian LGTs is usually rather large, of the order of the inverse of the glueball mass.
Thus, these results suggest that the value of $R_c$ computed in non-Abelian LGTs is mostly if not entirely due to the intrinsic width, with essentially no contribution from the quantum fluctuations of the EST.

Modeling the $\gamma_2$ dependence of our coefficients is beyond the scope of this work. In the spirit of our approach, we hope that our numerical results, reported in tables~\ref{tab:K2HT}, \ref{tab:K2LT}, \ref{tab:K4HT}, \ref{tab:K4LT} could be used to inspire and possibly benchmark future analytical studies of the string width in presence of the $\mathcal{K}^2$ correction.

\begin{figure}[h]
  \centering
  \includegraphics[scale=0.44,keepaspectratio=true]{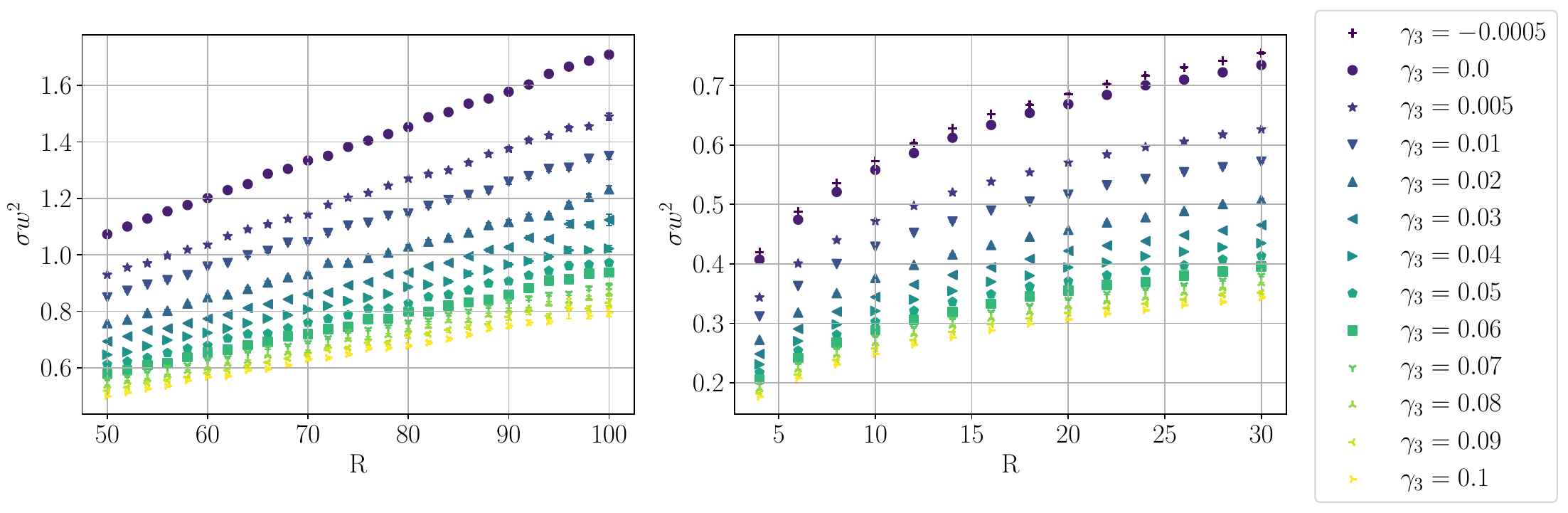}%
  \caption{Results for the width $\sigma w^2(R,L,\gamma_3)$ at high temperature ($R \gg L=20$, left panel) and at low temperature ($R \ll L=80$, right panel). The $\gamma_3=0$ case is equivalent to the Nambu-Got\={o} action with no $\mathcal{K}^4$ term.}
  \label{fig:wK4}
\end{figure}

We turn now to the study of the EST model described by eq.~\eqref{eq:NGK4action}, i.e. with inclusion of the $\mathcal{K}^4$ correction. We performed a similar analysis, both in the high- and low-temperature limits, choosing the same values of $\sigma$, $R$ and $L$ and $\gamma_3 \in [-0.0005,0.1]$. On a qualitative level, we found very similar results for the string width, which are fully reported in fig.~\ref{fig:wK4}. 

\begin{figure}[h]
  \centering
  \includegraphics[scale=0.6,keepaspectratio=true]{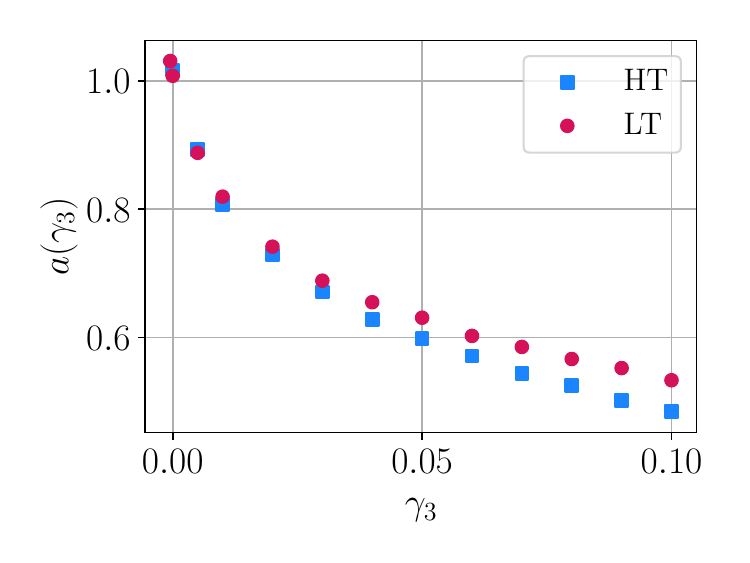}
  \includegraphics[scale=0.6,keepaspectratio=true]{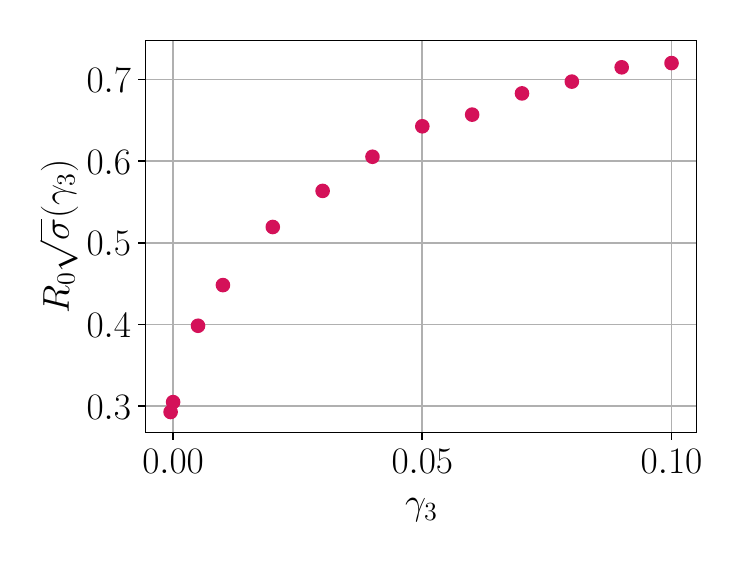}
  \caption{Results for $a^{(\HT)}(\gamma_3)$ and $a^{(\LT)}(\gamma_3)$ (left panel), and $R_c(\gamma_3)$ (right panel).}
  \label{fig:aRgamma4}
\end{figure}

As before, we fitted our data with the same functional forms discussed above, namely eqs.~\eqref{eq:wK2HT}, \eqref{eq:wK2LT} and \eqref{eq:K2R0b}, and we found the results reported in tables~\ref{tab:K4HT} and \ref{tab:K4LT} and plotted in figure~\ref{fig:aRgamma4} as a function of the $\gamma_3$ coefficient.



We conclude this section by stressing a few important consequences of our analysis.
The most interesting outcome of our simulations is that the effect of both the $\mathcal{K}^2$ or the $\mathcal{K}^4$ term on the string width is very strong. 
Conversely, in ordinary non-Abelian LGTs the behaviour of the string width is very well described by the pure Nambu-Got\={o} prediction: our results suggest that in the EST describing the gauge theories terms beyond the NG approximation should be either absent (as it is the case for the $\mathcal{K}^2$ term, due to the low energy universality) or very small (as it has been recently observed for the $\mathcal{K}^4$ term in refs.~\cite{Caristo:2021tbk, Baffigo:2023rin, Caselle:2024zoh}).
At the same time it is clear that the string width represents a powerful tool to detect the possible presence of a rigidity term in other gauge theories on the lattice, such as the trace-deformed LGTs or the three-dimensional $\mathrm{U}(1)$ model. 

Furthermore, it is particularly interesting to study the behaviour of the critical radius $R_c$ as a function of $\gamma_2$ and $\gamma_3$. 
Looking at figures~\ref{fig:aRgamma2} and \ref{fig:aRgamma4} we see that in both cases $R_c$ increases as a function of $\gamma_2$ or $\gamma_3$, but with different shapes. Modeling this behaviour might provide a tool to distinguish between the two types of corrections: in particular the critical radius seems to scale linearly with $\gamma_2$ in the $\mathcal{K}^2$ case. 
In this respect it is interesting to notice that a qualitatively similar behaviour of $R_c$ was recently observed in the three-dimensional $\mathrm{U}(1)$ model~\cite{Caselle:2016mqu}: in this setting confinement, thanks to the Polyakov solution~\cite{Polyakov:1976fu}, is indeed expected to be described by a rigid string. 
More precisely, in this model the rigidity term is expected to become more and more important as $\beta$ (the coupling constant of the model) increases and, remarkably, also $R_c$ is observed to increase as $\beta$ increases~\cite{Caselle:2016mqu}.

\subsection{Binder Cumulant}
\label{sec:binder}

We compute the Binder cumulant $U$ using eq.~\eqref{eq:binder} and we use it as a quantitative probe to understand whether the distribution of the field variable $\phi(\tau, R/2)$ (which contains info about the flux tube profile) is compatible with a Gaussian distribution (for which $U=0$) or not.

\subsubsection{The Nambu-Got\={o} action}

\begin{figure}[h]
  \centering
  \includegraphics[scale=0.5,keepaspectratio=true]{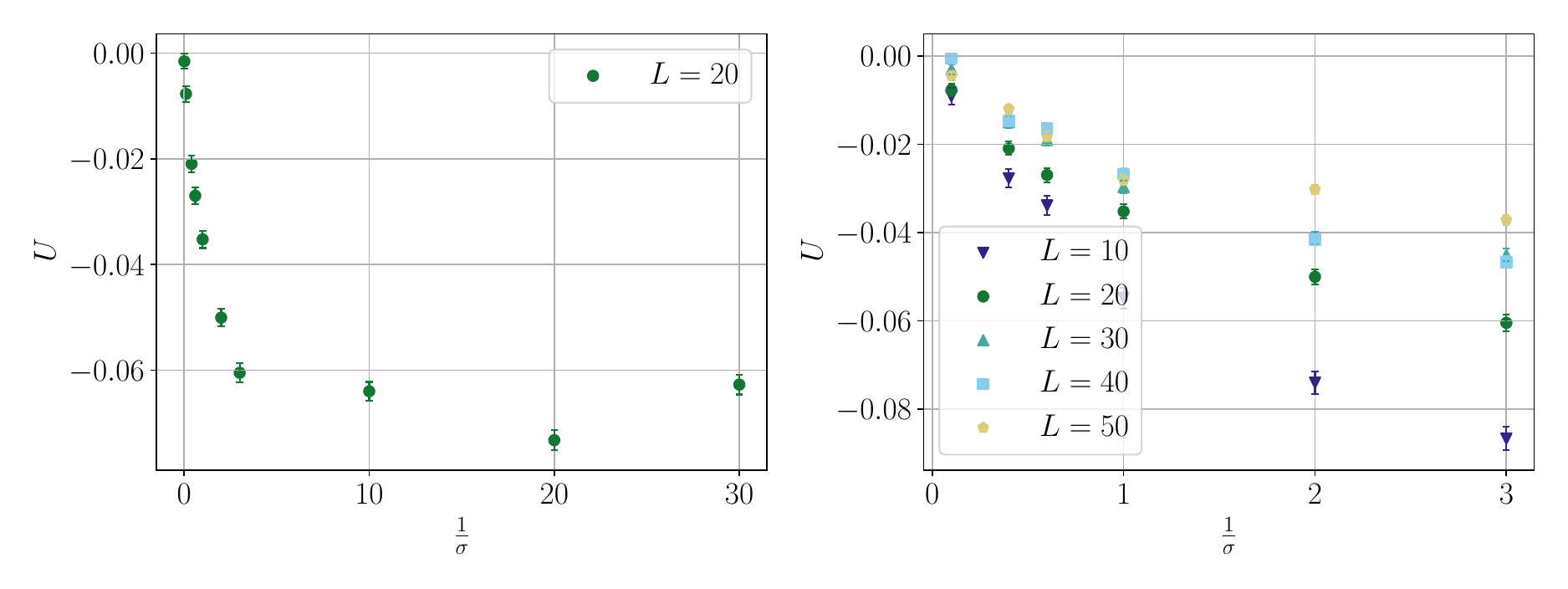}
  \caption{Results for the Binder cumulant $U$ as 
     a function of $1/\sigma$ for a $20\times 20$ lattice 
     (left panel) and for various lattice volumes $L\times L$ (right panel) 
     in the Nambu-Got{\=o} lattice model.}
  \label{fig:UNGsigma}
\end{figure}

In fig.~\ref{fig:UNGsigma} we report the results for $U$, for square lattices, in the case of the pure Nambu-Got\={o} action for decreasing values of $\sigma$ and various values of the volume $L\times L$ of the lattice. 

As expected, the cumulant is almost vanishing for large values of $\sigma$: in this regime we are very close to the Free Boson limit and we expect the profile of the flux tube to be Gaussian.
Interestingly, for decreasing values of the string tension, the cumulant becomes negative before reaching what appears to be a plateau for $\sigma \lesssim 1/3$, see left panel of fig.~\ref{fig:UNGsigma}.
This might suggest that the profile of the flux tube (i.e. our field variable $\phi$) is not Gaussian anymore as soon as the Free Boson approximation is not valid. 
However as we increase the lattice size $L$ this trend becomes weaker and weaker, see the right panel of fig.~\ref{fig:UNGsigma}, suggesting that a finite-size effect due to Dirichlet boundary conditions might be responsible for a nonzero $U$. 

\begin{figure}[h]
  \centering
  \includegraphics[scale=0.5,keepaspectratio=true]{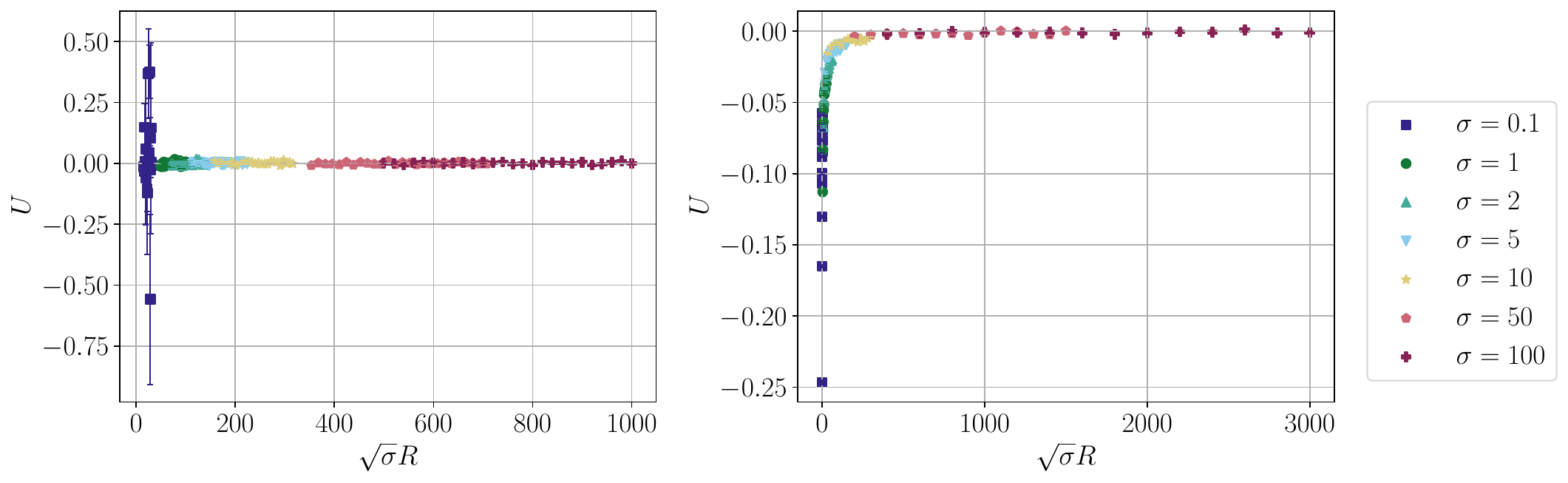}
  \caption{Results for the Binder cumulant $U$ as a function of $\sqrt{\sigma}R$ in the high-temperature limit ($R\gg L=8$, left plot) and in the low-temperature limit ($L=80\gg R$, right plot). Different colors represent different values of the string tension $\sigma$.}
  \label{fig:UNGsigmaLR}
\end{figure}

A possible way to test this issue is to study the NG model in an asymmetric setting, i.e. either in the high- or in the low-temperature limit, in which either $R$ or $L$ is very large. In fig.~\ref{fig:UNGsigmaLR} we report the results we obtained for the cumulant $U$ as a function of $\sqrt{\sigma}R$, for several combinations of $\sigma$ and $R$.

In the high temperature case we observe values of $U$ different from zero only for very small values of $\sqrt{\sigma} R$. These values have large errors and fluctuate randomly around zero. This behaviour agrees with the idea that the deviations are only due to the Dirichlet boundary conditions which are almost negligible in the high temperature limit.  
In the low temperature limit, i.e. when $L$ (the direction along which we fix Dirichlet boundary conditions) is large and the Dirichlet b.c. play a more important role, we see a systematic deviation of the Binder cumulant for small values of $\sqrt\sigma R$, which is exactly the region of parameters in which the effect of the Dirichlet boundary conditions may propagate more easily in the bulk.

\subsubsection{The Binder cumulant in presence of the $\mathcal{K}^2$ or the $\mathcal{K}^4$ terms}

\begin{figure}[h]
  \centering
  \includegraphics[scale=0.5,keepaspectratio=true]{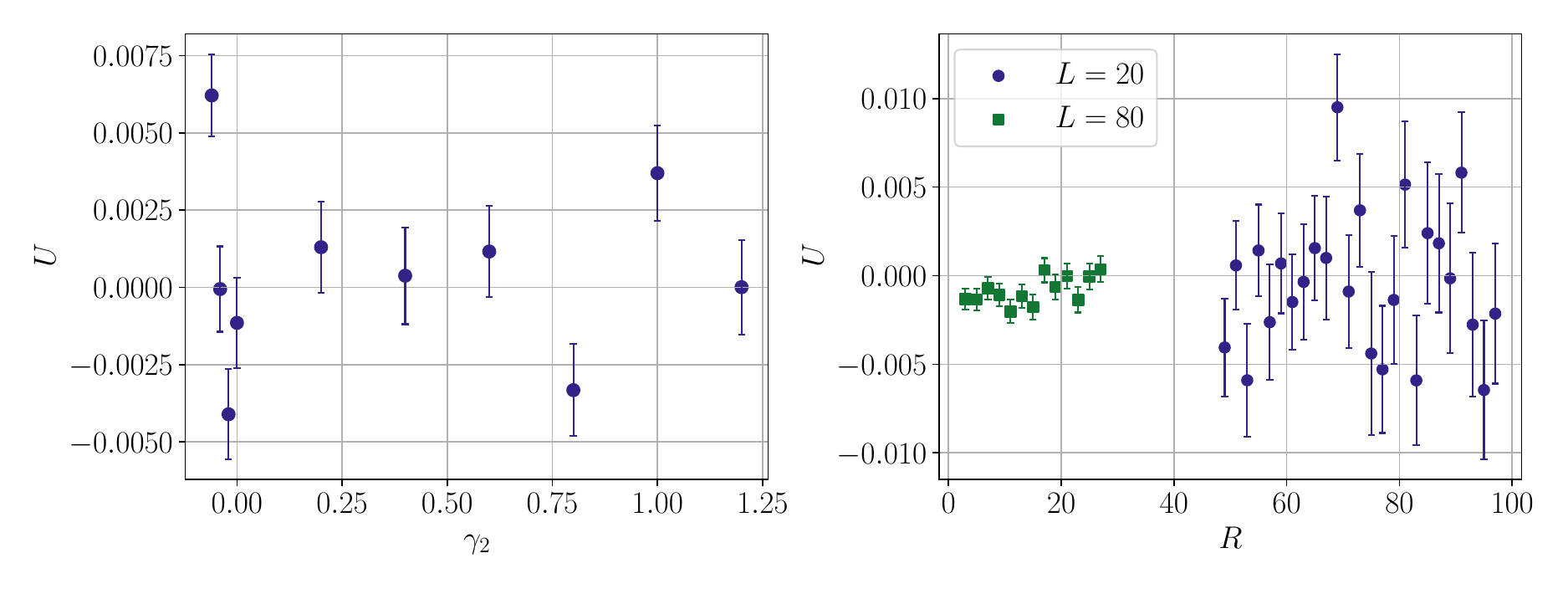}
  \caption{Results for the Binder cumulant $U$ in the NG model with the inclusion of the $\mathcal{K}^2$ term, as a function of $\gamma_2$ for a $20\times 20$ lattice (left panel) and as a function of $R$ for fixed $\gamma_2=0.02$ (right panel); in both plots $\sigma=100$.}
  \label{fig:UK2}
\end{figure}

\begin{figure}[h]
  \centering
  \includegraphics[scale=0.5,keepaspectratio=true]{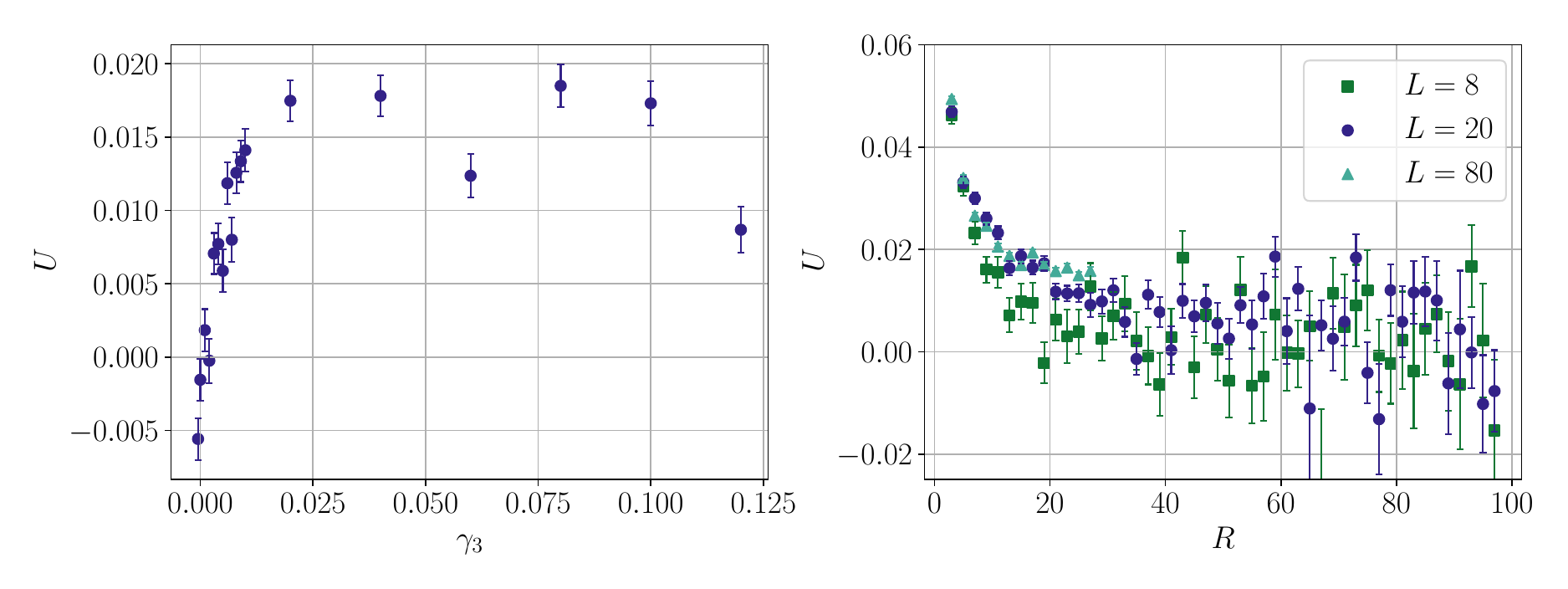}
  \caption{Results for the Binder cumulant $U$ in the NG model with the inclusion of the $\mathcal{K}^4$ term, as a function of $\gamma_3$ for a $20\times 20$ lattice (left panel) and as a function of $R$ for fixed $\gamma_3=0.02$ (right panel); in both plots $\sigma=100$.}
  \label{fig:UK4}
\end{figure}

We performed a similar analysis also for the actions which include the $\mathcal{K}^2$ or the $\mathcal{K}^4$ correction terms following a strategy similar to the one used in section~\ref{sec:resultsK2K4}. 
We set $\sigma=100$ and then we varied the coefficients $\gamma_2$ or $\gamma_3$ and the lattice sizes $L$ and $R$: results are reported in fig.~\ref{fig:UK2} and in fig.~\ref{fig:UK4} respectively. 
In both figures we report in the left panel a scan in the $\gamma_2$ or $\gamma_3$ coefficients for a $20\times 20$ lattice, while in the right panel a scan in $R$ for a fixed value of $\gamma_2$ or $\gamma_3$ and selected values of $L$.


In the case of the $\mathcal{K}^2$ action (the "rigid" string) we see essentially no deviations with respect to $U=0$. 
On the contrary, in the $\mathcal{K}^4$ case a small deviation from a vanishing Binder cumulant can be detected even for rather small values of $\gamma_3$, see fig.~\ref{fig:UK4}. 
Interestingly, this deviation is positive, i.e. it goes in the opposite direction with respect to what we observed in the NG case. 
As in the NG case its intensity decreases as $R$ increases and thus seems to be again a boundary effect. 

Let us close this section with a final important remark. In ordinary LGTs the Binder cumulant of the flux tube is typically always negative, much larger in magnitude than the values that we have found in the current work; furthermore, it shows a different dependence on $R$. 
Thus, the behaviour of the shape of the flux tube in LGTs is completely different from the ESTs that we study here~\cite{Verzichelli:2025cqc} and a quantitative description of its origin can be found in the intrinsic width of the confining flux tube that we mentioned above.

\section{Conclusions}

This paper is the natural completion of ref.~\cite{Caselle:2023mvh}: in that work, methods based on Normalizing Flows provided an efficient way to sample lattice EST models (like the ones discussed in this paper), which would be extremely challenging to tackle with standard Monte Carlo simulations otherwise. 
In this contribution, thanks to a novel flow-based architecture (the Stochastic Normalizing Flows described in section~\ref{sec:SNF}) we could overcome the limitations of the Continuous Normalizing Flow approach used in ref.~\cite{Caselle:2023mvh}. In particular, we could reach values of the string tension that allowed us to perform reliable numerical studies both of the Nambu-Got\={o} action and of models with additional terms proportional to a quadratic and a quartic power of the extrinsic curvature. A natural focus of our numerical effort was on the behaviour of the quantum width of the string, for which very few analytical results exist.

Our results have several relevant implication for LGTs studies of confinement. 
First of all, we have been able to measure the contribution to the string width of the terms included in the Nambu-Got\={o} string beyond the free bosonic approximation. In particular we have confirmed a conjecture on this behaviour proposed a few years ago in ref.~\cite{Caselle:2010zs}. 

Concerning terms beyond the Nambu-Got\={o} action proportional to powers of the extrinsic curvature, we have seen that they have a sizeable effect on the width: it is clear that this observable represents a natural probe for the identification of these higher order terms.
In particular the fact that in Monte Carlo simulations of ordinary non-Abelian LGTs the width of the flux tube follows rather precisely the expected Nambu-Goto behaviour is another proof of the fact that for these models the $\mathcal{K}^2$ term is absent (in agreement with the so called "low energy universality") and the $\mathcal{K}^4$ term has a very small coefficient~\cite{Caristo:2021tbk, Baffigo:2023rin, Caselle:2024zoh}.

Furthermore, we have shown that a powerful tool to detect the presence of a rigidity term in the action is the critical radius $R_c$ which, in presence of a $S_{\mathcal{K}^2}$ term in the action, grows linearly with the coefficient $\gamma_2$, see eq.~\eqref{eq:K2}.

Finally, we have shown that the corrections to the Gaussian behaviour are in general very small and show a dependence on $R$ perfectly compatible with boundary effects.
Crucially, the behaviour of the Binder cumulant in LGTs is pretty different from the one that we have observed for the EST models analyzed in this work. 
These results support a picture in which the shape of the flux tube in LGTs cannot be quantitatively described by the EST only, but requires the addition of a further contribution, the so-called "intrinsic width"~\cite{Caselle:2012rp}. In the EST language this contribution is associated to the critical radius $R_c$, which identifies the scale below which the EST description is not valid any more.

The main goal of this work was to show the feasibility of SNF simulations for a variety of effective string actions and the predictive power of numerical EST results. 
Our aim is not to solve these models, but to provide a reliable numerical framework that can work in synergy with analytical efforts in this direction. 
We think that the same methods could be applied to several other models, both in the context of more complicated EST constructions or in the study of other classes of effective theories. In particular, a natural candidate for further exploration is the extension of our model to two transverse degrees of freedom, which is the basic requirement for a EST description of four dimensional LGTs. In this context it would be interesting to study the so called "axionic string" which has been proposed in ref.~\cite{Dubovsky:2016cog} as a possible EST candidate for four dimensional LGTs.
Similarly, another candidate for further numerical work would be the Polchinski-Yang limit of the rigid string~\cite{Polchinski:1992ty}, for which an exact solution of the free energy exists, but no information of the width and shape is available.

\vskip 1.5cm
\noindent {\large {\bf Acknowledgments}}
\vskip 0.2cm
We thank Lorenzo Verzichelli for suggesting us the study of the Binder cumulant, and Andrea Bulgarelli, Marco Panero, Dario Panfalone for several insightful discussions. We acknowledge support from the SFT Scientific Initiative of INFN.
This work was partially supported by the Simons Foundation grant 994300 (Simons Collab-
oration on Confinement and QCD Strings) and by the European Union - Next Generation
EU, Mission 4 Component 1, CUPD53D23002970006, under the Italian PRIN “Progetti di
Ricerca di Rilevante Interesse Nazionale – Bando 2022” prot. 2022ZTPK4E.
\vskip 1cm

\appendix

\section{Fits of numerical results}

\begin{table}[h]
\centering
\begin{tabular}{|c |c c c c c|} 
 \hline
 $\sigma$ & $L$ & $a(L)$ & $b(L)$ & $c(L)$ & $\chisqred$\\ [0.5ex]
 \hline\hline
30 & 10 &  -516.87(3) & 19.8(2) & 0.78(8) & 0.63  \\
\hline
30 & 11 &  -568.17(4) & 22.5(3) & 0.6(1)& 0.89 \\
\hline
30 & 12 &  -619.60(5) & 24.4(4) & 0.6(1) & 1.17 \\
\hline
30 & 13 &  -670.95(4) & 27.1(4) & 0.5(1) & 0.90 \\
\hline
30 & 14 &  -722.38(4) & 29.3(4) &0.5(1) & 0.98 \\
\hline
30 & 15 &  -773.85(5) & 31.1(4)  & 0.5(1) & 1.05 \\
\hline\hline\hline
50 & 11 &  -1080.5(1) & 25.2(5) & 0.6(2) & 1.87 \\
\hline
50 & 12 &  -1178.26(7) & 27.6(3) & 0.6(1) & 0.65 \\ 
\hline
50 & 13 &  -1276.04(6) & 30.2(4) & 0.5(1) & 0.69 \\ 
\hline
50 & 14 &  -1373.93(7) & 32.4(3) & 0.6(1) & 0.58 \\ 
\hline
50 & 15 &  -1471.57(7) & 36.1(4) & 0.2(1) & 0.6 \\ 
 \hline\hline
\end{tabular}
\caption{Results for the coefficients of the fit of $\log Z$ of eq.~\eqref{eq:NGlnZ1}.}
\label{tab:logZNG}
\end{table}

\begin{table}[h]
\centering
\begin{tabular}{|c c c c|} 
 \hline
 $L$ & $f(L)$ & $g(L)$ & $\chisqred$\\ [0.5ex]
 \hline\hline
5  &  0.683(3)  &   2.6(1)  &   1.12  \\ 
\hline
6  &  0.562(3)   &   3.1(1)  &   0.85  \\ 
\hline
7  &  0.480(2) &   3.5(1)  &   1.78  \\ 
\hline
8  &  0.411(2)    &   3.79(7)  &   1.12  \\ 
\hline
9  &  0.372(1)  &   4.12(5)  &   0.57  \\ 
\hline
10  &  0.335(1)  &   4.33(6)  &   1.35  \\ 
\hline
11  &  0.3052(7)  &   4.55(3)  &   0.462  \\ 
\hline
12  &  0.2779(9)  &   4.81(4)  &   0.80  \\ 
\hline
13  &  0.2584(9)  &   4.89(4)  &   0.93  \\ 
\hline
14  &  0.2387(7)  &   5.17(3)  &   0.64  \\ 
\hline
15  &  0.2231(8)  &   5.29(4)  &   0.91  \\ 
\hline
16  &  0.2092(6)  &   5.46(3)  &   0.71  \\ 
\hline
17  &  0.1961(7)  &   5.60(3)  &   1.01  \\ 
\hline
18  &  0.1852(8)  &   5.76(4)  &   1.16  \\ 
\hline
19  &  0.1754(7)  &   5.88(3)  &   1.17  \\ 
\hline
20  &  0.1669(6)  &   5.98(3)  &   0.98  \\ 
 \hline\hline
\end{tabular}
\caption{Results for the coefficients of the fit of $\sigma w^2$ of eq.~\eqref{eq:width1}.}
\label{tab:widthNG}
\end{table}

\begin{table}[h]
\centering
\begin{tabular}{|c c c c|} 
 \hline
 $\gamma_2$ & $a^{(\HT)}(\gamma_2)$ & $b^{(\HT)}(\gamma_2)$ & $\chisqred$\\ [0.5ex]
 \hline\hline
-0.05 & 1.275(8) & 0.613(7) & 1.13  \\ 
\hline
-0.04 & 1.214(6) & 0.565(5) & 0.90 \\ 
\hline
-0.03 & 1.149(7) & 0.535(6) & 1.34 \\ 
\hline
-0.02 & 1.101(6)& 0.496(5) & 1.22 \\ 
\hline
-0.01 & 1.062(5) & 0.462(4) & 1.09 \\ 
\hline
0.0 & 1.013(4) & 0.445(4) & 1.01 \\ 
\hline
0.01 & 0.964(5) & 0.429(4) & 1.19 \\ 
\hline
0.02 & 0.931(5) & 0.408(4) & 1.14 \\ 
\hline
0.03 & 0.905(5) & 0.385(4) & 1.29 \\ 
\hline
0.04 & 0.867(4) & 0.375(3) & 0.80 \\ 
\hline
0.05 & 0.838(4) & 0.362(4) & 0.98 \\ 
\hline
0.06 & 0.816(4) & 0.344(4) & 1.05 \\ 
\hline
0.07 & 0.784(4) & 0.338(4) & 1.01 \\ 
\hline
0.08 & 0.760(5) & 0.327(4) & 1.33 \\ 
\hline
0.09 & 0.747(4) & 0.310(3) & 0.73 \\ 
\hline
0.1 & 0.723(5) & 0.304(4) & 1.31 \\ 
\hline\hline
\end{tabular}
\caption{Results for the coefficients of the fit of eq.~\eqref{eq:wK2HT} in the high-temperature limit.}
\label{tab:K2HT}
\end{table}

\begin{table}[h]
\centering
\begin{tabular}{|c c c c c|} 
 \hline
 $\gamma_2$ & $a^{(\LT)}(\gamma_2)$ & $b^{(\LT)}(\gamma_2)$ & $c^{(\LT)}(\gamma_2)$& $\chisqred$\\ [0.5ex]
 \hline\hline
-0.05 & 1.354(5) & 0.256(2) & 0.07(2) & 1.17 \\ 
\hline
-0.04 & 1.253(5) & 0.241(2) & 0.0(2) & 1.63 \\ 
\hline
-0.03 & 1.172(3) & 0.227(1) & -0.04(1) & 0.77 \\ 
\hline
-0.02 & 1.111(4) & 0.211(2) & -0.03(2) & 1.55 \\ 
\hline
-0.01 & 1.056(3) & 0.200(1) & -0.04(1) & 0.80 \\ 
\hline
0.0 & 1.008(3) & 0.188(1) & -0.04(1) & 1.16 \\ 
\hline
0.01 & 0.966(4) & 0.178(2) & -0.048(2) & 2.29 \\ 
\hline
0.02 & 0.928(2) & 0.170(1) & -0.04(1) & 0.74 \\ 
\hline
0.03 & 0.902(2) & 0.1579(9) & -0.015(8) & 0.59 \\ 
\hline
0.04 & 0.871(2) & 0.1508(7) & -0.021(7) & 0.46 \\ 
\hline
0.05 & 0.841(3) & 0.145(1) & -0.03(1) & 1.28 \\ 
\hline
0.06 & 0.817(2) & 0.138(1) & -0.02(1) & 1.10 \\ 
\hline
0.07 & 0.795(2) & 0.1314(8) & -0.010(7) & 0.62 \\ 
\hline
0.08 & 0.770(3) & 0.127(2) & -0.02(1) & 2.25 \\ 
\hline
0.09 & 0.758(3) & 0.119(1) & 0.01(1) & 1.66 \\ 
\hline
0.1 & 0.738(3) & 0.114(1) & 0.02(1) & 1.35 \\ 
\hline\hline
\end{tabular}
\caption{Results for the coefficients of the fit of eq.~\eqref{eq:wK2LT} in the low-temperature limit.} 
\label{tab:K2LT}
\end{table}

\begin{table}[h]
\centering
\begin{tabular}{|c c c c|} 
 \hline
 $\gamma_3$ & $a^{(\HT)}(\gamma_3)$ & $b^{(\HT)}(\gamma_3)$ & $\chisqred$\\ [0.5ex]
 \hline\hline
0.0 & 1.016(4) & 0.440(4) & 0.82 \\ 
\hline
0.005 & 0.893(5) & 0.371(4) & 0.79 \\ 
\hline
0.01 & 0.807(6) & 0.348(5) & 0.88 \\ 
\hline
0.02 & 0.729(7) & 0.299(5) & 1.14 \\ 
\hline
0.03 & 0.671(6) & 0.272(5) & 1.35 \\ 
\hline
0.04 & 0.628(4) & 0.253(4) & 0.69 \\ 
\hline
0.05 & 0.598(5) & 0.234(4) & 1.10 \\ 
\hline
0.06 & 0.571(5) & 0.222(4) & 1.09 \\ 
\hline
0.07 & 0.544(4) & 0.217(3) & 0.72 \\ 
\hline
0.08 & 0.526(4) & 0.206(3) & 0.77 \\ 
\hline
0.09 & 0.502(4) & 0.205(4) & 0.57 \\ 
\hline
0.1 & 0.485(5) & 0.198(4) & 0.92 \\ 
\hline\hline
\end{tabular}
\caption{Results for the coefficients of the fit of eq.~\eqref{eq:wK2HT} in the high-temperature limit for a model including both the NG and $\mathcal{K}^4$ terms in the action.}
\label{tab:K4HT}
\end{table}

\begin{table}[h]
\centering
\begin{tabular}{|c c c c c|} 
 \hline
 $\gamma_3$ & $a^{(\LT)}(\gamma_3)$ & $b^{(\LT)}(\gamma_3)$ & $c^{(\LT)}(\gamma_3)$& $\chisqred$\\ [0.5ex]
 \hline\hline
-0.0005 & 1.031(4) & 0.195(2) & -0.05(2) & 1.08 \\
\hline
0.0 & 1.008(4) & 0.189(2) & -0.06(2) & 1.35 \\
\hline
0.005 & 0.889(3) & 0.146(1) & 0.02(1) & 0.71 \\
\hline
0.01 & 0.820(4) & 0.128(2) & 0.05(2) & 1.02 \\
\hline
0.02 & 0.742(4) &  0.104(2) & 0.7(1) & 0.91 \\
\hline
0.03 & 0.689(5) & 0.091(2) & 0.09(2) &  2.03  \\
\hline
0.04 & 0.655(4) & 0.080(2) & 0.11(2) & 1.43 \\
\hline
0.05 & 0.631(3) & 0.070(1) & 0.12(1) & 0.90 \\ 
\hline
0.06 & 0.603(4) & 0.067(2) & 0.10(2) & 1.66 \\ 
\hline
0.07 & 0.586(3) & 0.061(1) & 0.121(9) & 0.58 \\ 
\hline
0.08 & 0.567(3)  & 0.057(1) & 0.107(9) & 0.69 \\ 
\hline
0.09 & 0.553(3) & 0.053(1) & 0.11(1) & 0.91 \\ 
\hline
0.10 & 0.534(3) & 0.052(1) & 0.10(1) &  0.77 \\ 
 \hline\hline
\end{tabular}
\caption{Results for the coefficients of the fit of eq.~\eqref{eq:wK2LT} in the low-temperature limit for a model including both the NG and $\mathcal{K}^4$ terms in the action.} 
\label{tab:K4LT}
\end{table}
\bibliography{biblio}

\end{document}